\documentclass[twocolumn,showpacs,preprintnumbers, nofootnbib, prd,superscriptaddress,10pt]{revtex4-1}
\usepackage{physics}
\usepackage{graphicx}
\usepackage{mathtools}
\usepackage{amssymb}

\begin{document}

\title{The Role of Entropy in the Evolutionary Quantization of the Isotropic Universe}%

\author{Andrea Campolongo}%
\email{andreacampolongo95@gmail.com}
\affiliation{Physics Departement, Sapienza University of Rome, Piazzale Aldo Moro 5, 00185 Rome, Italy}

\author{Giovanni Montani}
\email{giovanni.montani@enea.it}
\affiliation{Physics Departement, Sapienza University of Rome, Piazzale Aldo Moro 5, 00185 Rome, Italy}
\affiliation{ENEA, Fusion and Nuclear Safety Departement, C.R. Frascati, via Enrico Fermi 45 (00044) Frascati (RM), Italy}

$\date{}$
\begin{abstract}{In this paper, we analyze the dynamics of an isotropic closed Universe in the presence of a cosmological constant term and we compare its behavior in the standard Wheeler-DeWitt equation approach with the one when a Lagrangian fluid is considered in the spirit of the Kuchar-Brown paradigm.\\
		In particular, we compare the tunnelling of the Universe from the classically forbidden region to the allowed one, showing that considering a time evolution deeply influences the nature of the model. In fact, we show that in the presence of the Lagrangian fluid, the cosmological singularity is restored both in the classical and the quantum regime. However, in the quantum regime the singularity is probabilistically suppressed for some energy eigenvalues and in the case the latter is equal to zero, one recovers the standard WDW case.\\
		Finally, we introduce a cut-off physics feature in the Minisuperspace by considering a Polymer quantum mechanical approach, mainly limiting our attention to the semi-classical dynamics (the quantum treatment is inhibited by the non-local nature of the Hamiltonian operator). We show that the singularity is again removed, like in the fluid-free model, and a bouncing cosmology emerges so that the present model could mimic a cyclic cosmology.}%
\end{abstract}
\maketitle
\section{Introduction}

One of the most puzzling questions of the canonical quantization of gravity is the so-called \emph{frozen formalism}, i.e. 
the absence of an external time parameter for the quantum dynamics 
of the $3$-metric field \cite{primordial,1,rovelli,Isham92}.
Furthermore, while the Loop Quantum Gravity implementation 
to cosmology leads to the existence of a Big-Bounce 
\cite{Ashtekar}, the Wheeler DeWitt equation, associated to a metric approach, seems, in general unable to remove the cosmological singularity on a quantum level \cite{blyth,benini}. The main reason for the 
quantum survival of the cosmological singularity consists of 
the time-like character of the Universe volume in the Wheeler Super-space 
\cite{DeWitt,Gravitation}.
In fact, the Wheeler-DeWitt equation resembles a functional 
Klein-Gordon equation, for which the $3$-metric determinant 
behaves like an internal clock. As a result, all the values 
of this quantity are available to the dynamics, including its 
vanishing character, associated to the singularity. 
This feature is deeply altered in Loop Quantum Gravity 
since the $3$-volume acquires a discrete spectrum. \\
A non-singular Universe can be easily obtained in the 
Einsteinian dynamics if we include a positive cosmological 
constant into the evolution of an isotropic Universe 
\cite{primordial}.
In this respect, a very intriguing \emph{no boundary proposal} has been formulated in \cite{7}, see also \cite{6}, which argues the possibility of a tunnelling effect from the classically forbidden vanishing scale factor to  a finite volume region, living also on a classical level  (for a simple canonical representation of this idea, 
see \cite{8}). \\
A delicate question concerning the point of view that  the quantum primordial Universe could have undergone a tunnelling procedure comes from the absence of time in the canonical quantum dynamics, which makes this notion heuristic. In addition, it seems also in contradiction that, for more general models, the Universe volume is itself a time-like variable, while the real degrees of freedom are identified in the Universe anisotropy \cite{primordial}. \\
A viable methodology for introducing a good time variable in quantum gravity is the ``Kuchar-Brown" method, proposed in \cite{kuchar-brown}, see also \cite{thiemann}. In \cite{cianfrani}, this procedure has been implemented to a Lagrangian fluid representation, as presented in \cite{20}, by the analysis in \cite{cianfrani}, where the fluid entropy  has been promoted to be the proper clock of a Schr{\"o}edinger-like 
functional equation. \\
Here, we apply the results of the study mentioned above \cite{cianfrani} towards a quantum picture for the early isotropic Universe. 
We consider a closed Robertson-Walker geometry whose dynamics 
includes a cosmological constant term and a Schutz Lagrangian fluid. \\
We first study the classical dynamics of this cosmological model, 
for which the presence of a singularity is restored because the 
Lagrangian fluid takes the morphology of radiation-like 
component of the Universe. Then, we analyze the quantum behavior, 
studying the configurational properties of the scale factor, especially for what concerns the possibility of a tunnelling effect through the potential barrier.\\ 
We show that the cosmological singularity is, in general, 
present in such a model, although it seems to be probabilistically 
suppressed for wave packets associated to eigenstates having 
the ``energy-like" quantum number smaller than the 
potential peak. 
Thus, the typical configuration for a Universe tunnelling scenario 
becomes a singular quantum cosmology, as soon as the notion 
of time is properly restored for the classical and quantum dynamics. \\
In order to remove the obtained singularity in the 
considered cosmological model  we introduce a notion of 
cut-off physics in the Minisuperspace via the Polymer 
quantum mechanics approach \cite{corichi,mantero,barca}. 
\\Actually, the implementation of a semi-classical Polymer approach  shows how the singularity is removed and a Big-Bounce emerges also in the presence of a radiation-like fluid, playing the role of a clock. The pure quantum treatment of this revised scenario 
is not viable, due to the non-local character of the associated 
Hamiltonian operator. \\
The paper is structured as follows. In Section 2, we introduce the Schutz formalism for a relativistic perfect fluid and we derive its Hamiltonian theory coupled to gravity. Then the Wheeler-DeWitt approach seen in \cite{7,6} will be discussed in Section 3 just before studying both the classical and quantum dynamics that arise when the Kuchar-Brown method is applied to a Universe in which there is a Schutz fluid (Section 4).
Finally, the Polymer representation will be introduced and its dynamics will be approached.

\section{The Schutz fluid as a viable clock in Quantum Gravity}
In this paper, the problem of time is approached through a canonical quantization. The direct implementation of the Hamiltonian constraints leads to a well-known non-evolutive theory;
\begin{equation}
 \mathcal{H}_G = \int d^3x(NH_G + N^aH^G_a)
\end{equation} 
where $H_G$ and $H^G_a$ are respectevely the ``Super-hamilto- nian" and the ``Super-momentum" constraints and $N$ and $N^a$ are the Lapse function and the Shift vector of the ADM splitting.\\
One of the many attempts to solve this problem is considering a theory coupled to gravity and searching for a time parameter out of the inner variables of the theory.\\
In this paper, it will be followed an approach in which is added a Schutz fluid to gravity  \cite{20,40} and the entropy per baryon will be chosen as time variable.\\
A Schutz fluid is a relativistic perfect fluid whose four velocity $U^{\mu}$ is written as a combination of five scalar potential\\
\begin{equation}
\label{4-velocity}
 U_{\nu} =\frac{1}{\mu}(\phi,_{\nu}+\alpha \beta,_{\nu}+\theta S,_{\nu}) :=\frac{1}{\mu}v_{\nu} , 
\end{equation}
where $S$ is the entropy per baryon and $\mu = (\rho+p)/\rho_0$ is the specific enthalpy of the fluid.
Such a choice is considered for the fact that the Schutz fluid can better approximate the primordial thermal bath. The latter is not properly characterized by an ultra-relativistic fluid but, it could be better described by a Schutz fluid whose equation of state $p=\alpha\rho$ is such that $\alpha=\alpha(\mu,S)$ is thermodinamical-variable dependent.\\
The fluid's equations of motion are derived as usual from a variational principle varying the lagrangian density
\begin{equation}
\label{lag}
\resizebox{.9\hsize}{!}{$\mathcal{L}^{ADM}_F=\sqrt{-g}p=N\sqrt{-^3g}\rho_0\Bigl(\sqrt{(v_n)^2-v_av^a}-TS\Bigr)$}
\end{equation}\\
with respect to the fields that compose the four-velocity.
In the fluid's lagrangian $\rho_0$, is the density of the rest mass and $\emph{T}$ is the temperature.\\
The coupled theory is given by deriving the fluid's Hamiltonian in the ADM formalism as usually done for the gravitational counterpart. \\
For what regards the Schutz Hamiltonian, it has to be derived in a Dirac manner \cite{50} given that from $(\ref{lag})$ one obtains a number of second class constraints $\phi_{\alpha}=0$. Solving those constraints, as prescribed by the Dirac theory, leads to the fluid's Hamiltonian \cite{cianfrani}
\begin{equation}
H_{F}= N  \Bigl(\sqrt{(\pi^2-\rho_0^2h)V}+\rho_0\sqrt{h}TS \Bigr)+N^a\pi v_a,
\end{equation}
where $\pi$ is the momentum conjugated to $\phi$, $\sqrt{-^3g}=\sqrt{h}$ and $V=v_{\mu}v^{\mu}$.\\
The coupling with gravity comes natural and does not change the constrained nature of the theory.
The secondary constraints will appear as usual in the total Hamiltonian as the multiplication of functionals with the lapse function and the shift vector:
\begin{equation}
\mathcal{H}_{F+G} = \int d^3x(NH_{F+G} + N^aH^{F+G}_a)
\end{equation} 
where \\
\begin{equation}
\label{Htot}
 H_{F+G} = \sqrt{(\pi^2-\rho_0^2h)V}+\rho_0\sqrt{h}TS + H^G 
\end{equation}
\begin{equation}
 H^{F+G}_a = \pi v_a + H^G_a.
\end{equation}
Being the constraint nature unaffected, a canonical quantization of the ``Super-Hamiltonian" constraint leads to a time-less Schr{\"o}edinger equation.\\
A time evolution can be established through the ``Kuchar- Brown" method \cite{kuchar-brown} using the Schutz fluid as a clock.\\
The method consists in choosing an inner variable of the theory as a new time variable and expliciting the dynamics with respect to the chosen one. In doing so, one gets an equivalent ``Super-Hamiltonian" constraint whose canonical quantization leads to an evolutive Sch$\scalebox{0.75}[1.0]{-}$ r{\"o}edinger equation.\\
This is achieved by solving the ``Super-momentum" constraint with respect to the momentum relative to the chosen variable and inserting it into the ``Super- Hamiltonian" constraint.\\ The latter reads as a request of a gradient-free Hamiltonian (a good time parameter is always the same everywhere) and leads to an equation for $\pi$ 
\begin{equation}
 \pi-\tilde{h}=0.
\end{equation}
Finally it has to be checked that $\{\tilde{h},\tilde{h}\}=0$ strongly; this way the new Hamiltonian $\tilde{h}$ can be interpreted as the generator of time translations and the four-diffeomor- phism invariance is preserved.\\
If the Kuchar-Brown method is applied to the case in analysis one gets
\begin{equation}
  \tilde{h}=  \sqrt{\frac{d\rho_0^2h}{\Xi^2-d}}
\end{equation}
where $ \Xi=\sqrt{h}\rho_0ST+ H^G$ and $d=H^G_aH^G_bq^{ab}$.\\
The role of entropy as a time variable emerges when the particular comoving reference frame is chosen.\\
In this setting the conjugated momentum reduces to $\pi=-\sqrt{q}\rho_0$ and the Hamiltonian constraints become\\
\begin{equation}
\label{secondary}
 \Xi = \sqrt{h}\rho_0ST+H^G=0
\end{equation}
\begin{equation}
 \Xi_a = H^G_a=0.\\
\end{equation}
Recalling the relation between $\pi$ and the momentum conjugated to S $(\phi_{\alpha}\rightarrow p_S-\theta\pi=0)$ and applying the discussed method, one gets
\begin{equation}
\label{eq1}
 Sp_S =  \frac{\theta H^G}{T} = \bar{h}.
\end{equation}
The canonical quantization of the latter leads to a \\ Schr{\"o}edinger equation for the Hamiltonian $\bar{h}$ where the time parameter is given by the logarithm of the entropy per baryon, $\log S$.\\
The study of the latter will be performed in Section 4.

   \section{The Wheeler-De Witt approach}
   
 The Kuchar-Brown method allows to restore a time evolution for the Hamiltonian of gravity. However, it is possible to perform a direct canonical quantization of the ``Super-Hamiltonian" constraint. In doing so, the ``Wheeler-De Witt" equation $(\hat{H}\psi=0)$ is given.
 \begin{figure}
 	{\includegraphics[scale=0.75]{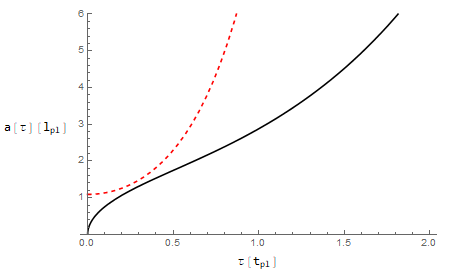}}
 	\caption{The figure shows the comparison between the solutions of the Friedman equation with (continuous line) or without (red dotted line) the presence of a Schutz fluid. Both the cases describe a closed Universe in which there is a cosmological constant ($\frac{\Lambda}{3} \propto 10^{-1}E_{pl}$) }
 	\label{classical}
 \end{figure}
 The attempt of studying what could be the dynamics descending from such an approach has been already attempted \cite{7,6} but, in this scenario, one has to think how to deal with a time-less Schr{\"o}edinger equation.\\
 This approach will be briefly presented in order to compare it with the one that will be shown in the following sections.\\ 
 Firstly, an isotropic and homogeneous Universe, described by the Robertson-Walker metric, is considered (in this paper we will use natural units, $c=\hbar=1$),
 \begin{equation}
 \resizebox{.9\hsize}{!}{$ds^2=dt^2-a(t)^2\Bigl(\frac{dr^2}{1-kr^2}+r^2d\delta^2+r^2sen^2\delta d\phi^2\Bigr)$}
 \end{equation}
 where $a(t)$ is the scale factor and $k$ is the curvature parameter.\\
 In both the Wheeler-De Witt case and the studied one, a spatially closed Universe ($k>0$), in which there is a Cosmological constant, will be taken into account. \\
 Its classical dynamics is the one of a bouncing Universe $(Fig. {\ref{classical}})$; a Universe characterized by a minimal value for the scale factor $a_{MIN}(\tau)$ at a given time. The latter is given by the solution of the Friedmann equation
 \begin{equation}
 \label{fri}
  \tilde{H}^2= \Bigl(\frac{\dot{a}}{a}\Bigr)^2=\Bigr(\frac{\Lambda}{3}\Bigr)-\frac{k}{a^2}\,
 \end{equation}
 for the Universe in analysis. The Friedmann equation is derived considering the Hamilton equations associated to the Hamiltonian
 \begin{equation}
 \label{hamcosmo}
  H^G_ {FRW}=-\frac{G}{3\pi}\frac{p_a^2}{a}-\frac{3\pi}{4G}(ka-\frac{\Lambda}{3}a^3).
 \end{equation}
 The corresponding WDW equation for a closed FRW Universe, when a factor ordering $j=0$ $(p_a^2=-\frac{\partial^2}{\partial a^2})$ is given, is 
 \begin{equation}
 \label{eqschrohh}
  \Bigl( \frac{\partial^2}{\partial a^2}-\frac{9\pi^2}{4G^2}(ka^2-\frac{\Lambda}{3}a^4)\Bigr)\psi(a) =0.
 \end{equation}
 If the latter is expressed via dimensionless quantities at Planck scales such as $a_0=(\Lambda/3)^{-1/2}=G^{1/2}$, the unbounded potential 
 \begin{equation}
  U(a)=  \frac{9\pi^2a_0^2}{4G^2}\Bigl [\Bigl(\frac{a}{a_0}\Bigr)^2-\Bigl(\frac{a}{a_0}\Bigr)^4\Bigr ]
 \end{equation}
 will determine two different regions: there is a classically not-allowed one $(0<a<a_0)$ and a classically allowed one $(a>a_0)$.
 \begin{figure}[b]
 	{\includegraphics[scale=0.68]{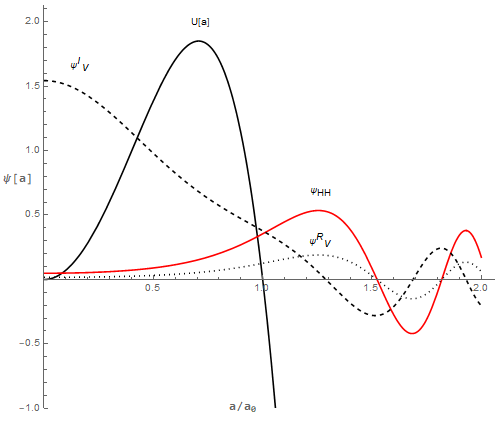}}
 	\caption{The figure shows the potential $U(a)/3$ (black curve), the wave functions for Hartle-Hawking (red curve) and for Vilenkin boundary conditions (dashed and dotted curves). The real and the imaginary parts of Vilenkin's wave functions are indicated with R and l respectively. The Hartle-Hawking wave function is real. The figure displays the solutions to the WDW equation when the factor ordering $j=1$ is chosen. This choice will not affect the semi-classical dynamics \cite{8}.}
 	\label{pot_vil}
 \end{figure}
 Despite the under barrier region, there is always a probability different from zero that a particle could emerge from the barrier into the allowed region.
 This model has been firstly discussed in a path integral approach{\cite{7}}, and the tunnelling probability has been computed.
 This probability can be obtained through the WKB approximation and it is given by
 \begin{equation}
     P = e^{-2\int \sqrt{U(x)}} \approx e^{-S_E}
 \end{equation}
 where $S_E=-\frac{3\pi}{\Lambda G}$ is the Euclidean action.
 The WKB method can be used when one is searching for the eigenfunctions of the WDW approach as done for the probability. Different solutions can be derived when different boundary conditions are considered.
 In doing so, the eigenfunctions of $(\ref{eqschrohh})$ are 
 \begin{equation}
 \label{wkb1}
 \psi^{(1)}_{\pm}(a)= e^{\pm i \int_{a_0} ^{a} |p(a')|da' \mp i\frac{\pi}{4}}
 \end{equation}
 in the classically allowed region and, in the classically forbidden one,
 \begin{equation}
 \label{wkb2}
 \psi^{{2}}_{\pm}(a)= e^{\pm \int_a ^{a_0} |p(a')|da'},
 \end{equation}
 where $p(a')=\sqrt{-U(a)}$.
 If an outgoing wave solution is chosen \cite{6}, one has
 \begin{equation}
 \label{wkb3} 
 \psi_{OUT}(a>a_0)=\psi^{(1)}_-(a)
 \end{equation}
 \begin{equation}
 \psi_{OUT}(a<a_0) = \psi^{(2)}_+(a)-\frac{i}{2}\psi^{(2)}_-(a)
 \end{equation}
 whereas, if an expanding and contracting Universe is chosen \cite{7}
 \begin{equation}
 \label{wkb4} 
 \psi_{E+C}(a>a_0)=\psi^{(1)}_+(a)+\psi^{(1)}_-(a)
 \end{equation}
 \begin{equation}
 \psi_{E+C}(a<a_0)=\psi^{(2)}_-(a).\\
 \end{equation}
 Those solutions can be combined through the WKB connection formula in order to obtain a sole eigenfunction valid everywhere.\\
 Whatever solution is considered, both of them show that in the classical forbidden region one has a non-null eigenfunction and so, a non-null tunnelling probability.
 However, even though the tunnelling probability obtained through the WKB method is non null, the time-less nature of the theory does not allow a proper ``tunnelling effect"  as we know from standard quantum mechanics. In fact, the absence of a time means the absence of a time ordering of the events. 
 So, the introduction of a proper time parameter becomes a key factor in describing the quantum dynamics.
  
   \section{A Specific model for the Isotropic Universe dynamics}
   The Kuchar-Brown method allows to establish a time parameter, avoiding the problems that a time-less approach carries with. At the same time, the introduction of a fluid, representing the primordial thermal bath, changes both the quantum and the classical dynamics.\\
   From the classical perspective, adding a Schutz fluid is equivalent to adding an ultra-relativistic component to the Hamiltonian first, and then to the Friedmann equation.\\
   In fact, the new Hamiltonian is
   \begin{equation}
   \label{hamfrw+fluido}
   H_{G+F}=\sqrt{h}\rho_0TS-\frac{\chi}{24\pi^2}\frac{p_a^2}{a}-\frac{3\pi}{4G}ka+2\pi^2\rho a^3=0
   \end{equation}
   and the first term can be considered as an energy density $\rho(a)=\rho_0a^{-3(1+\omega)}$ with $\omega=\frac{1}{3}$ as anticipated.\\
   The Friedmann equation will be then
   \begin{equation}
   \label{fri}
   \tilde{H}^2=\Bigl(\frac{\dot{a}}{a}\Bigr)^2=\Bigl(\frac{8\pi G}{3}\frac{1}{a^4}-\frac{k}{a^2}+\frac{\Lambda}{3}\Bigr),
   \end{equation}
   whose solutions now represent a singular Universe $(Fig. \ref{classical})$.
   The latter could not be true when a quantum analysis is approached. One could say that the dynamics will be in accordance with the classical one only if a well peaked wave packet can be achieved in the spirit of the Ehrenfest theorem.\\
   Passing to the proper quantum analysis, the equation $(\ref{eq1})$ becomes 
   \begin{equation}
   \label{27}
    -i \frac{\partial{\psi(x,\tau)}}{\partial{\tau}} = \frac{\theta}{T}\mathcal{H}^G_{FRW}\psi(x,\tau)
   \end{equation}
   where an FRW Universe is considered. Here $\tau = \log S$  and $\frac{\theta}{T}$ is a function to be determinated.\\
   This ratio could be fixed; in fact from the Schutz' model is known that $\theta$  is one of the potentials which is connected to a physical quantity via $\frac{\partial{\theta}}{\partial{t}} = T$, where the derivative is taken with respect to the proper time.\\
   It is then possible to compute this integral and get a function $\theta=\theta(a) $ by ``guessing" which is the functional dependence of $\dot{a}(t)$.\\ For the case in analysis one will take $\dot{a}(t)$ as given by a Friedmanian dynamics ruled by radiation as it was confirmed from the classical dynamics  shown above. \\
   If  one does so, the integral obtained is:
   \begin{equation}
   \label{apunto}
   \resizebox{.9\hsize}{!}{$\theta (a) = \int\frac{\partial{\theta}}{\partial{\tau}}d\tau = \int\frac{T(a)}{\dot{a}} da =  \int\frac{T(a)}{aH}da = \sqrt{\frac{3}{8\pi G}}a$}
   \end{equation}
   and then the ratio will be $\frac{\theta}{T} = \sqrt{\frac{3}{8\pi G}}a^2$.\\
   In order to study the equation $(\ref{27})$, one may take the wave function's time dependence to be given by
   \begin{equation}
    \psi(a,\tau)=\psi(a)e^{-iE\tau};
   \end{equation}
   this way one gets the Schr{\"o}edinger equation:
   \begin{equation}
   \resizebox{.87\hsize}{!}{${a}\frac{\partial^2}{\partial a^2}\psi(a,\tau)  =\Bigr( \frac{9 \pi^2}{4G^2} (ka^3 - \frac{\Lambda}{3}a^5)- \sqrt{24\pi^3}E \Bigr)\psi(a,\tau)$} .
   \end{equation}
   Now, if the equation is expressed via dimensionless quantities at Planck  scales as done for the WDW case and a canonical transformation is performed, one gets 
   \begin{equation}
   \label{scroe}
   \frac{\partial^2}{\partial x^2}\psi(x) = (U(x)-\tilde{E})\psi(x)
   \end{equation}
   where $\tilde{E}= \sqrt{24\pi^3} E$, $x=(\frac{a}{a_0})^2$ and 
   \begin{equation}
   \label{potential2}
    U(a)= \frac{9\pi a_0^4}{4G^2}\Bigl[\Bigl(\frac{a}{a_0}\Bigr)^3-\Bigl(\frac{a}{a_0}\Bigr)^5\Bigr].
   \end{equation}
   
   \subsection{Quantum FRW analysis}
   Before going deep into the quantum dynamics, one recalls that the introduction of a time parameter gives the chance of properly describing a tunnelling effect.
   \begin{figure}[!h]
   	\includegraphics[scale=0.6]{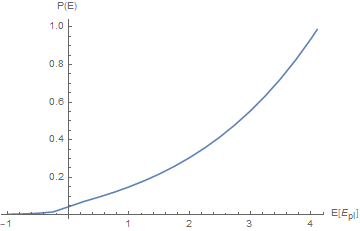}
   	\centering
   	\caption{The Figure shows the energy dependence of the probability. As expected for a tunnelling process in a potential $(\ref{potential2})$ ($P(E) \rightarrow 1 , E \rightarrow U_{MAX}$) meanwhile ($P(E) \rightarrow 0 , E \rightarrow -\infty$)}
   	\label{fig:prob}
   \end{figure}
   The identification of the logarithm of the entropy with a time allows to define the tunnelling probability with respect to the eigenvalue $\tilde{E}$
   \begin{equation}
    P \approx e^{-2\int \sqrt{U(x)-\tilde{E}}}.
   \end{equation}
   As it is showed by ($Fig.\ref{fig:prob}$) the probability $P(E)$ goes to $P(E)\rightarrow 1$ when the energy approaches the potential maximum $(V_{MAX} \approx 4.1 E_{pl})$, meanwhile $P(E) \rightarrow 0$ when the energy parameter is such that $E \rightarrow -\infty$.\\ 
   In addition, it is worth noting that the probability $P(E=0)=e^{-\pi}$ it's the same one could have got from a standard WDW approach \cite{7,6} confirming the coherence of the model and its generality.\\
   Furthermore, the tunnelling probability exponentially decreases with the decrease of the eigenvalue $\tilde{E}$. \\
   It is then possible to find the eigenfunctions of  $(\ref{scroe})$ and their time evolution.
   \begin{figure*}
   	\includegraphics[scale=0.9]{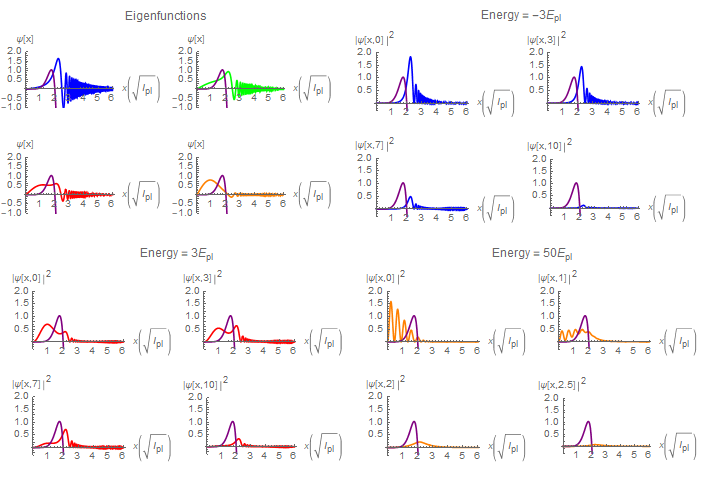}
   	\centering
   	\caption{The panel shows the the numerical analysis done for the quantum FRW model. The first figure shows the solution of the Schr{\"o}edinger equation $(\ref{scroe})$ taken for different energy eigenvalues: $E= -3E_{pl}$, $E= E_{pl}$, $E= 3E_{pl}$ and $E\gg V_{Max}= 50E_{pl}$. Then, in the other three figures, the evolution of a wave packet peaked at different energies has been studied ($\sigma=1.5$), plotting the wave packet at different times. The potential plotted is $U(x)/4$; this is done in order to help the understanding of the wave packet's behavior}
   	\label{fig:E=-3}
   \end{figure*}
   The latter is done by building a wave packet peaked at the eigenvalue chosen and studying its time evolution with respect to the changes in the eigenvalue.\\
   Both those processes are done numerically.
   In order to underline the different features of the eigenfunctions, different solutions of $(\ref{scroe})$ will be compared.\\
   From those functional form one will be able to anticipate the features that the wave packet's time evolution will have for each energy that will be treated next.\\
   At first glance, it is clear their common high oscillatory feature; all the Schr{\"o}edinger equation's solutions $(Fig.\ref{fig:E=-3}. a)$ have this characteristic on the right side of the potential. This is due to the potential form; in fact, whatever the energy $\tilde{E}^*$ is, there will be a point where it can not be comparable with the potential which rapidly decreases to infinity. So, the solution will be an eigenfunction which increases its oscillatory frequencies as the ``x-coordinate" increases.\\
   What is interesting is their different behavior when the energy eigenvalue changes. In fact, as it can be seen, the more $\tilde{E}$ increases, the more  the eigenfunction is peaked on the left-side of the potential.\\
   This feature will be a key one when the wave packet analysis will be performed. In fact, it can be anticipated that for energies low enough it will not be possible to peak a wave packet on the left side of the potential; this would be possible only for some energy eigenvalues such as $\tilde{E}=3  E_{pl}$.\\
   The solutions' time evolution is done, as anticipated, by studying the evolution of a well peaked wave packet.
   The latter is given by
   \begin{equation}
    \psi_{\tilde{E}^*}(x,\tau)=\int d\tilde{E}g(\tilde{E},\tilde{E}^*)\psi(x,\tilde{E})e^{-iE\tau}
   \end{equation}
   where $\psi(x,\tilde{E})$ is the numerical solution of $(\ref{scroe})$ for a given energy $\tilde{E}$ and $g(\tilde{E},\tilde{E}^*)$ is a gaussian distribution centered in $\tilde{E}^*$
   \begin{equation}
    g(E,E^*)=\frac{1}{\sqrt{2\pi}\sigma}e^{-\frac{(E-E^*)^2}{2\sigma^2}}.
   \end{equation}
   Starting from negative eigenvalues $(Fig.\ref{fig:E=-3}.b)$, one finds that the more $\tilde{E}^*$  is negative the more the wave packet is peaked on the right side of the potential and the probability of finding the Universe near the singularity exponentially decreases.\\
   So, the more the eigenvalue increases, the more  probability of finding the Universe near the singularity exponentially increases. From a certain value $E^*$ onwards,  $|\psi(x,\tau)|^2$ will be represented by a wave packet initially peaked on the left-potential.
   However, as long as the energy eigenvalue increases, the tunnelling probability does so ($Fig.\ref{fig:prob}$).\\
   So, even though at the initial time the Universe wave function is left-peaked, as time goes by the Universe will tunnel through the potential barrier.
   Once the potential barrier is overcome, for all the energy eigenvalues, the wave packet will rapidly spread and the Universe will expand.\\
   So, the introduction of the time parameter gives a proper ``tunnelling effect" characterization.
   The Kuchar-Brown method allows to solve the Super-Hamiltonian constraint and describe the quantum dynamics via the choice of a Schutz fluid as a clock. \\
   However, having introduced a fluid changes both of the classical and quantum Universe dynamics.
   In fact, the quantum dynamics describes a ``Bounce-less" Universe in which the singularity is probabilistically suppressed.\\
   This, even though it is a great achievement for the model's sake of completeness, does not  assure the existence of a singularity-free Universe. In addition to that, for the case $E\gg V_{MAX}$  $(Fig.\ref{fig:E=-3}.d)$ one has the dynamics of a free wave packet which can be peaked near the singularity.\\
   This issue can be solved by having a natural process that lets the Universe having a bouncing dynamics and so, one may suggest to take into account a Polymer dynamics scenario.
   \section{Polymer quantum mechanics}
  Let us introduce then the Polymer quantum mechanics which is a different mechanical scheme from the standard Schr{\"o}edinger one.
  It is an independent quantization procedure which has been introduced for its analogy with Loop Quantum Cosmology (LQC) which is given by the possibility of deriving both the Loop's Hilbert space and the semi-classical dynamics.
  The Polymer mechanics is a particular representation in which the Stone-Von Neumann theorem  is not satisfied, providing a unitarily inequivalent representation and, as a consequence, different physical predictions \cite{corichi}.\\
  In order to appreciate the freedom of choosing the representation, one considers the Weyl Algebra given by the exponentiation of the operators position and momentum, that will be indicated with $\hat{q}$ and $\hat{p}$ respectively.
  The construction of a ``Fock space" can be done defining the complex structure $J$ which acts on the phase space $\Gamma= \mathbb{R}^2$ such that $J^2=-1$.\\
  The Hilbert space can be obtained from the Weyl albebra via the GNS Construction (Gel’fand-Naimark-Segal).
  What is subtle is that considering the Weyl algebra there are some choices of $J$ for which the Stone-Von Neumann theorem  is not satisfied and thus, are not equivalent to the Schr{\"o}edinger one.
  In those cases the Polymer representation arises.\\
  The result of such an approach is the inability of properly defining both the $\hat{q}$ and $\hat{p}$ operators.\\
In order to study the Polymer representation, one can consider an Hilbert space $\mathcal{H}$, some abstract kets $\ket{\mu}$ with $\mu \, \epsilon \, \mathbb{R}$ and some subsets defined by $\mu_i \, \epsilon \,  \mathbb{R}$ with $i=1,...,N$.
Then, if those kets are taken to be orthonormal $\braket{\mu}{\nu}=\delta_{\mu \nu}$, one can define a Hilbert space $\mathcal{H}_{poly}$ on which two different operators act, a label and a displacement one
  \begin{equation}
    \hat{\epsilon}\ket{\mu}=\mu\ket{\mu} 
  \end{equation}
  \begin{equation}
    \hat{s}(\lambda)\ket{\mu}=\ket{\mu+\lambda}.
  \end{equation}
  The shift operator $s(\lambda)$ will be discontinuous since all the kets are orthonormal and so, it cannot be obtained from the exponentiation of any Hermitian operator.\\
  In order to connect this abstract representation to physical systems and physical operators one may consider a Hamiltonian system with canonical variables $q$ and $p$. If the momentum polarization is chosen, the fundamental states are 
  \begin{equation}
  \psi_{\mu}(p)=\braket{p}{\mu}=e^{ip\mu}.
  \end{equation}
  So, according to what has been previously said the label operator $\hat{\epsilon}$ will be identified with the position operator $\hat{q}$
  \begin{equation}
   \hat{q}\psi_{\mu}\coloneqq-i\partial_p\psi_{\mu}= \mu\psi_{\mu}
  \end{equation}
  whereas the shift operator role will be taken by the multiplicative operator $V(\lambda)$
  \begin{equation}
   \hat{V}(\lambda)\psi_{\mu}\coloneqq e^{i\lambda p}e^{ip\mu}=\psi_{\mu+\lambda}
  \end{equation}
  from which becomes clear that $\hat{p}$ cannot be taken as the generator of translations.
 In this case can be shown that the Hilbert space for a generic representation is given by
 \begin{equation}
 \qquad \mathcal{H}_{poly}=L^2(\mathbb{R}_B,d{\mu}_{H})
 \end{equation}
 where $d{\mu}_{H}$ is the Haar measure and $\mathbb{R}_B$ is the Bohr compactification of the real line.\\
  Things do not change if the position polarization\\ $\psi(q)=\braket{q}{\psi}$ is chosen. In this case it can be shown that the wave functions are Kroneker deltas, the translation operator is discontinuous and so the momentum operator can not be well defined.\\
  In this scenario it can be demonstrated that, the Hilbert space is $\mathcal{H}_{poly}=L^2(\mathbb{R}_d,d{\mu}_{c})$, where this time  $\mathbb{R}_d$ is the real axis with discrete topology meanwhile $d{\mu}_{c}$ is the counting measure \cite{corichi}.\\
  In order to overcome the problems stemming from the definition of both $\hat{q}$ and $\hat{p}$, one introduces the graph $\gamma_{\mu_0}=\{q \in \mathbb{R} \> |q=n\mu_0 \> \forall n \in \mathbb{Z}\}$, where $\mu_0$ is the scale introduced in the Polymer scheme.\\
  The Hilbert space that arises taking into account the graph $\gamma_{\mu_0}$ is such that $\mathcal{H}_{\gamma_{\mu_0}} \subset \mathcal{H}_{poly}$ and it will contain the states 
  \begin{equation}
    \ket{\psi}= \sum_n b_n\ket{\mu_n}
  \end{equation}
  where $\mu_n=n\mu_0$ and $\sum_n|b_n|^2<\infty$.\\
  The displacement operator will change according to the fact that now the shift is set by the lattice spacing $\mu_0$, leading to the result
  \begin{equation}
   \hat{V}(\mu_0)\ket{\mu_n}=\ket{\mu_0+\mu_n}=\ket{\mu_{n+1}}
  \end{equation}
  Knowing how the shift operator acts one can build a regulated operator $\hat{p}_{\mu_0}$.\\
  Considering the case when $p\ll1/{\mu_0}$ the momentum can be approximated by
  \begin{equation}
  \label{p2}
   p \approx \frac{1}{\mu_0}\sin(\mu_0p) = \frac{1}{2i\mu_0}\Bigl(e^{i\mu_0p}-e^{-i\mu_0p}\Bigr)
  \end{equation}
  and then the regulated operator will be\\
  \begin{align}
     \hat{p}_{\mu_0}\ket{\mu_n}=&\frac{1}{2i\mu_0}(\hat{V}(\mu_0)-\hat{V}(-\mu_0))\ket{\mu_n} =\\
  &=\frac{1}{2i\mu_0}(\ket{\mu_{n+1}}-\ket{\mu_{n-1}}). 
  \end{align}\\
  Different approximations are possible when the regulated squared momentum operator is considered.\\
  One may consider to compose the operator $\hat{p}_{\mu_0}$ with itself. This leads to an operator which shifts to two steps the states in the graph
  \begin{equation}
  \hat{p}^2_{\mu_0}\ket{\mu_n}=\frac{1}{4\mu^2_0}(2-\hat{V}(2\mu_0)-\hat{V}(-2\mu_0))\ket{\mu_n}
  \end{equation}
  and so 
  \begin{equation}
  \hat{p}^2\approx \frac{1}{\mu^2_0}\sin^2(p\mu_0).
  \end{equation}
  On the other hand, if an operator which shifts only once is considered one has
  \begin{equation}
  \hat{p}^2_{\mu_0}\ket{\mu_n}=\frac{1}{\mu^2_0}(2-\hat{V}(\mu_0)-\hat{V}(-\mu_0))\ket{\mu_n}
  \end{equation}
  and so 
  \begin{equation}
  \hat{p}^2\approx \frac{2}{\mu^2_0}\Bigr(1-\cos(p\mu_0)\Bigr).
  \end{equation}
  With these considerations one can take into account the Hamiltonian operator which lives on the space $\mathcal{H}_{\gamma_{\mu_0}}$
  \begin{equation}
   H_{\gamma_{\mu_0}}=\frac{1}{2m}\hat{p}^2_{\mu_0}+\hat{V}(\hat{q})
  \end{equation}
  where $\hat{V}(\hat{q})$ is the potential.\\
  The dynamics that comes from the Hamiltonian can be studied in the momentum polarization where the momentum operator acts like a multiplicative operator 
  \begin{equation}
  \hat{p}^2\psi(p)\approx \frac{1}{\mu^2_0}sin^2(p\mu_0)\psi(p)
  \end{equation}
  whereas the $\hat{q}$ operator is represented by the derivative operator
  \begin{equation}
  \hat{q}\psi_{p}\psi(p)=i\partial_p\psi(p).
  \end{equation}
  This ends the process of building the polymer dynamics. \\
  What is interesting is how to recover the physical Hilbert space $\mathcal{H}_S=L^2(\mathbb{R},dp)$ from the Polymer one $\mathcal{H}_{\gamma_0}$.\\
  It is not possible to embed $\mathcal{H}_S$ in $\mathcal{H}_{\gamma_0}$, in fact $\mathcal{H}_S$ cannot be obtained by dividing $\mu_0$ into smaller and smaller intervals.\\
  However one can try to approximate a continuous wave function with a function defined on $\mathcal{H}_{\gamma_0}$. Once the real line $\mathbb{R}$ is decomposed in $n$ intervals which defines a scale $C_k$, the wave function will be represented by a function constant in each of those intervals.
 As a result, at any given scale $C_k$ one will have an approximated  kinetic term $(\ref{p2})$ of the Hamiltonian operator and so, a number of  effective theories which will be related by coarse-graining maps \cite{corichi}.
   
   \subsection{Semi-classical Polymer dynamics}
\begin{figure}
   	\includegraphics[scale=0.6]{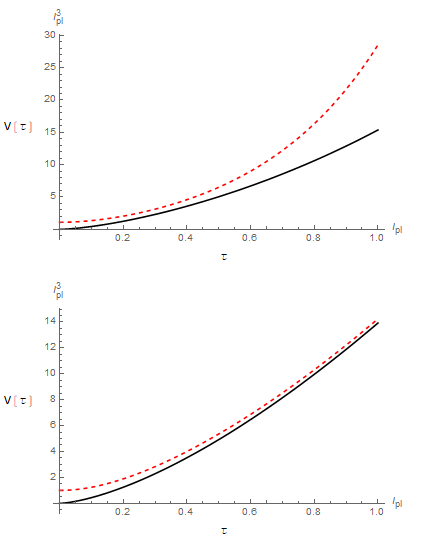}
   	\centering
   	\caption{The Figure shows the classical dynamics behavior for a flat Universe (solid line) with Cosmological constant $\Lambda \neq 0$ versus the semi-classical Polymer one (red-dashed line). Two different cases are displayed the one for $\Lambda = 10^{-1} E_{pl}$ (first picture), and the one for $\Lambda =  10^{-4} E_{pl}$. As it can be seen, in the second case the semi-classical dynamics tends to the classical one}
   	\label{fig:lambdasemipoly}
   \end{figure}

   It must be said that in this Section some features of the Polymer quantum mechanics will be given to the standard Friedmann dynamics without developing the full quantum theory.  This will be in fact done at a semi-classical level, in the spirit of the Ehrenfest theorem, searching for the main changes  in the classical dynamics when the Polymer quantum mechanics is considered.\\
   The new Hamiltonian obtained starting from $(\ref{hamcosmo})$, considering the momentum approximation, will be
   \begin{equation}
   \label{semipoly}
   \resizebox{.9\hsize}{!}{$\mathcal{H}^{poly}_{FRW}=-\frac{3G}{\pi}\frac{\sin^2(\mu_0P_V)}{\mu_0^2}V+2\pi^2\rho_{TOT}(V) V-\frac{3\pi}{4G}kV^{\frac{1}{3}}=0$}
   \end{equation}
   where $\rho_{TOT}$ takes into account the fluid's presence.
   In order to derive the latter, the canonical transformation $a^3\rightarrow V$ has been performed; one chooses the volume $V$ as the preferred Polymer variable because with this choice the critical density $\rho_c$ will be scale factor independent \cite{mantero}.
   The Polymer approximation will then change the Friedmann equation 
   \begin{equation}
   \label{qq}
   \quad \tilde{H}^2=\Bigl(\frac{1}{3}\frac{\dot{V}}{V}\Bigr)^2=\frac{\chi}{3}\rho\Bigl(1-\frac{\rho}{\rho_{\mu}}\Bigr),  \qquad \rho_{\mu}=\frac{B}{4\pi^2}\frac{1}{\mu_0^2}
   \end{equation}
   (where $B=\frac{3\chi}{4\pi^2}$) adding the term $\Bigl(1-\frac{\rho}{\rho_{\mu}}\Bigr)$ to the standard Friedmann equation.\\
   This equation represents the semiclassical equation at scale $\mu_0$. It is worth noting that for $\mu_0\rightarrow 0$, and therefore $\rho_{\mu}\rightarrow \infty$, one obtains the standard Friedmann equation.
   In order to see how the Polymer approximation changes the semiclassical dynamics (seen in Section 4), one uses $(\ref{semipoly})$ and  expresses it via dimensionless quantities at Planck scales. This way one obtains
   \begin{equation}
   \label{qq}
    H^2=\Bigl(\frac{1}{3}\frac{\dot{\tilde{V}}}{\tilde{V}}\Bigr)^2=\frac{\chi}{3}Q\Bigl(1-\frac{Q}{Q_{\mu}}\Bigr)
   \end{equation}
   where $\tilde{V}(l_{pl}^3)$ is the dimensionless volume, Q is
   \begin{equation}
    Q=\Bigl(\frac{\rho_{TOT}(\tilde{V})}{\rho_{pl}}-\frac{3}{8 \pi}\frac{k}{\tilde{V}^{\frac{2}{3}}}\Bigr)
   \end{equation}
   and 
   \begin{equation}
   \label{qu}
    Q_{\mu}=\frac{3}{2\pi^3 \mu_0^2}\frac{G}{\rho_{pl}},
   \end{equation}
   which, given the fact that $\mu_0=[E]^{-3}$, $G=[E]^{-2}$ and $\rho_{pl}=[E]^{4}$, is dimensionless. Moreover the simple request $Q_{\mu}=1$, which is done in order to simplify the calculation automatically fixes the Polymer parameter
   \begin{equation}
    \mu_0(l_{pl}^3)=\sqrt{\frac{3}{2\pi^3}} \approx 0.22.
   \end{equation}
   and so the Polymer lattice parameter $L_{poly}=\sqrt[3]{\mu_0}l_{pl}\approx 0.60l_{pl}$ \cite{barca}.\\
 The momentum approximation acts on the Friedmann equation introducing a cut-off density and so, a cut off on the volume.
In fact, since there is a finite value of the density $\rho = \rho_{\mu}$ for which  $\tilde{H}^2$ can be zero, the volume's time dependence will have a critical point, hence the solution of the modified Friedmann equation will represent a Big-Bounce Universe $(Fig.\ref{fig:lambdasemipoly})$.
   
   \subsection{Quantum Polymer dynamics}
\begin{figure}[h]
   	\includegraphics[scale=0.8]{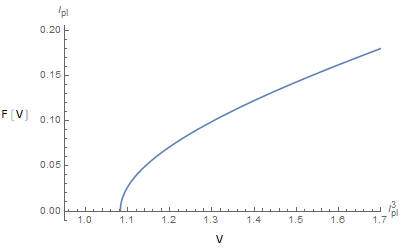}
   	\centering
   	\caption{The Figure shows the function  $F(V)=\frac{\theta(V)}{T(V)}$ obtained through a numerical integration}
   	\label{fig:theta}
   \end{figure}

\begin{figure}[h]
   	\includegraphics[scale=0.75]{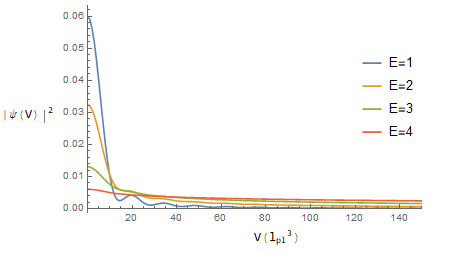}
   	\centering
   	\caption{The Figure shows the function  $|\psi(V)|^2$ solution of $\ref{aaa}$ for different energy eigenvalues. The more the energy grows the more  $|\psi(V)|^2$ is broadened}
   	\label{fig:penultimo}
   \end{figure}

   Finally, it is possible to approach the Polymer quantum mechanics analysis of the model. Starting from $(\ref{hamfrw+fluido})$ and recalling the value of the fluid's momenta, it is possible to obtain the equation
   \begin{equation}
   \label{eqqpoly}
    p_sS=\frac{\theta}{T}\Bigl(-\frac{B}{2}P_V^2V+2\pi^2\rho_{\Lambda}V-\frac{3\pi}{4G}kV^{\frac{1}{3}}\Bigr).
   \end{equation}
   In Section 4, in order to study the quantum dynamics, the ratio $\theta/T$  was set and then a canonical transformation was performed.\\ 
   Following the same procedure, this ratio can be obtained through dynamical considerations, where now $\dot{a}(t)$ is no longer given by the standard Friedman equation but from the Polymer modified one $(\ref{qq})$.
   In doing so, one gets
   \begin{equation}
    \theta(a)=\int\frac{d\theta}{da}\frac{1}{\dot{a}}da=\int T(a)\frac{1}{\dot{a}}da
   \end{equation}
   which, if one considers that the dynamical polymer variable is the volume V, becomes
   \begin{equation}
   \label{thetaV}
    \int\frac{1}{\dot{V}}V^{-\frac{1}{3}}dV=\theta(V)
   \end{equation}
   where the relation $\dot{a}=\frac{1}{3}\Bigl(\frac{\dot{V}}{V}\Bigr)V^{\frac{1}{3}}$ has been used.
   Then considering that $\dot{V}$ is given by the semi-classical Polymer dynamics of a flat Universe with $\Lambda =10^{-1} E_{pl}\neq 0$
   \begin{equation}
    \Bigl(\frac{1}{3}\frac{\dot{\tilde{V}}}{\tilde{V}}\Bigr)^2=\frac{8\pi}{3}(Q_F+Q_{\Lambda})(1-\frac{Q_F+Q_{\Lambda}}{Q_{\mu}})
   \end{equation}
   is then possible to numerically integrate $(\ref{thetaV})$ and obtain the function $F(V)=\frac{\theta(V)}{T(v)}$ $(Fig.\ref{fig:theta})$.\\
   It is clear that $F(V=V_{min})=0$ by construction, in fact
   \begin{equation}
    F(V)=\frac{\int_{V_{min}}^V\frac{1}{\dot{V}}V^{-\frac{1}{3}}dV}{T(V)}
   \end{equation}
   but, what arises is that F(V) has infinite derivative in $V=V_{min}$.\\
   So, it is not possible to fix the ratio $\frac{\theta}{T}$ through the semiclassical dynamics as done for the non-Polymer case (Section 4). \\
   It is then possible to try to put a cut-off on $F(V)$, in this way one could take into account the bouncing feature of the dynamics. One of the possible choices is trying to put a cut-off on the temperature $T=T_{MAX}$ this way,
   \begin{equation}
    F(V)=\frac{\int_{0}^tT(\tau)d\tau}{T(t)} \approx\frac{ T_{MAX}t_{pl}}{T_{MAX}} = t_{pl}.
   \end{equation}
   So, taking the equation $(\ref{eqqpoly})$ and performing the substitution $(\ref{p2})$ one will have
   \begin{equation}
   \resizebox{.87\hsize}{!}{$p_sS=\frac{\theta}{T}\Bigl(-B\bigl(\frac{\sin^2(\mu_0 P_V)}{\mu_0^2}\bigr)V+2\pi^2\rho_{\Lambda}V-\frac{3\pi}{4G}kV^{\frac{1}{3}}\Bigr)$}.
   \end{equation}
   After having performed a canonical Polymer quantization and expressed it via dimensionless quantities at Planck scales, it becomes
   \begin{equation}
   \label{aaa}
   \resizebox{.9\hsize}{!}{$E\psi(P_V)= \Bigl(2\pi^2Q_{\Lambda}-4\pi^2sin^2(\mu_0 P_V)\Bigr)\frac{\partial}{\partial P_V}\psi(P_V)$}
   \end{equation}
   where the relation $\mu_0=\sqrt{\frac{3}{2\pi^3}} l_{pl}^{-3}$ was used and it was considered a flat Universe.\\
   This equation has been solved for different energies and then its Fourier transform has been studied (Figure: $\ref{fig:penultimo}$).
   As it can be seen, it is not possible to have a well peaked wave packet in a volume's value other than $V=0$. In addition it is not even possible to talk about a localized Universe at all. \\
   Therefore, in such a model the Universe will not have a Big-Bounce dynamics but also, it will not have a semi-classical dynamics at all.\\
   This is not what one should expect from the Polymer quantum approach of a model which semiclassically has a Bounce dynamics. The reason why this happens is because one has chosen a particular value for $F(V)$ and each different choice leads to a different Polymer dynamics.\\ In fact the natural way of determining $F(V)$ gives a singular function in $V=V_{MIN}$ $(Fig.\ref{fig:theta})$ while using a cut-off leads to a non-Bounce dynamics.\\
   However this does not mean that the Polymer quantum dynamics implementation is not able to reproduce its semiclassical feature. This means that the $F(V)$ choice must be worked out and the possibility of working with non-local operators has to be taken into account. In fact the path followed in determining the ratio  $\theta/T$ led to the numeric integration $(Fig.\ref{fig:theta})$
   \begin{equation}
     \frac{\theta}{T}(V)\propto \sqrt{V-V_{MIN}}
   \end{equation}
   which operational implementation leads to a non-local theory and its Taylor expansion is not useful as said before.\\
   This point has to be clarified in order to derive a coherent dynamic and it will be objective of future works.
\section{Concluding remarks}
We analyzed a revised version of  the simplest canonical formulation for the so-called no-boundary proposal, regularizing it by the 
inclusion of a matter clock in the spirit of the Kuchar-Brown prescriptions. \\
We consider the cosmological implementation of a previous 
analysis in which a Lagrangian Schutz fluid is properly 
addressed to indentify its specific entropy as a viable clock for a quantum gravity theory. 
In particular, we considered an isotropic closed Universe, in the presence of a cosmological constant, adding a Schutz fluid to get an evolutionary quantization of its dynamics. \\
We demonstrated that the Schutz fluid behaves, on a classical level, as a radiation-like component of the Universe and therefore it restores a singularity in the past, absent when only the cosmological 
constant term is present.\\
Our quantum analysis demonstrates that such a singularity is still present on a quantum level, but its probabilistic weight strongly depends on the range of the energy-like eigenvalue we are considering. The probability of finding the scale factor in the classically forbidden region is strongly suppressed when the energy-like parameter is significantly smaller than the potential maximum or it is negative. \\
The merit of this analysis consists in showing how the regularization of a tunnelling of the Universe by including time into the quantum dynamics produces a removal of the classical bounce due to the cosmological constant alone. 
This result suggests that conjectures based on a frozen quantum dynamics are not necessary valid once matter is included to make the dynamics evolutionary. However, its introduction clarifies the meaning of the tunnelling process, restoring the concept of ``before'' and ``after''. \\
We also demonstrated that a bouncing cosmology is immediately recognized when cut-off physics features are introduced via a Polymer quantum mechanical approach. Our study has precise validity in the semi-classical regime only, since a pure quantum analysis in the Polymer framework is inhibited by the non-local nature of the resulting Hamiltonian operator. \\
It must be regarded as an interesting topic for further investigations to discuss the present paradigm in more general dynamical contexts, like the homogeneous Bianchi Universes. The aim of such a generalization of the present study could be clarifying if the tunnelling procedure of the Universe is systematically affected by the nature of the considered clock and therefore conjectures like the no-boundary proposal must be properly addressed and revised in an evolutionary framework.


\begin{thebibliography}{1}
	
	\bibitem{primordial} G. Montani, M. V. Battisti, R. Benini, G. Imponente, Primordial Cosmology, World Scientific, (2011)
	
	\bibitem{1} F. Cianfrani, O. M. Lecian, M. Lulli, G. Montani, Canonical Quantum Gravity, World Scientific, (2014)
	
	\bibitem{rovelli} C. Rovelli, Classical Quantum Gravity 8, 1613 (1991).
	
	\bibitem{Isham92} Isham, C., Canonical quantum gravity and the problem of time, (1992). arXiv:gr-qc/9210011 
	
	\bibitem{Ashtekar} A. Ashtekar, T. Pawlowski, and P. Singh, Phys. Rev. Lett. 96, 141301, (2006).
	
	\bibitem{blyth} Blyth, W. and Isham, C., Phys.Rev. D11, pp. 768–778, (1975). doi:10.1103/PhysRevD.11.768
	
	\bibitem{benini} Montani, G., Battisti, M. V., Benini, R. and Imponente, G., Int.J.Mod.Phys. A23, pp.2353–2503, (2008). doi:10.1142/S0217751X08040275, arXiv:0712.3008
	
	\bibitem{DeWitt} DeWitt, B. S., Physical Review 160, pp. 1113–1148, (1967).
	
	\bibitem{Gravitation} Misner, C. W., Thorne, K. S. and Wheeler, J. A., Gravitation, Freeman, W. H. and C., San Francisco, (1973).
	
	\bibitem{7} J. b: Hartle and S.W. Hawking, Phys. Rev,  D28, 2960 (1983).
	
	\bibitem{6} A. Vilenkin,  Phys. Rev, D37,888 (1988).
	
	\bibitem{8} E. W. Kolb and  M. S. Turner, The Early Universe, Addison-Wesly, (1988).
	
	\bibitem{kuchar-brown} Brown, J. D. and Kuchař, K. V., Phys.Rev. D51, pp. 5600–5629, (1995). doi: 10.1103/PhysRevD.51.5600, arXiv:gr-qc/9409001
	
	\bibitem{thiemann} T. Thiemann, "Solving the Problem of Time in General Relativity and Cosmology with Phantoms and k-Essene," ArXiv Astrophysis e-prints,
	July 2006.
	
	\bibitem{cianfrani} F. Cianfrani, G. Montani, S. Zonetti, 2009. arXiv:0807.3281v2 
	
	\bibitem{20} B. F. Schutz, Jr., Phys, Rev. D 2, 2762, (1970).
	
	\bibitem{corichi} A. Corichi, T. Vukasinac and J. A. Zapata, Class. Quant. Grav. 24, 1495, (2007).
	
	\bibitem{mantero} G. Montani, C. Mantero, F. Bombacigno, F. Cianfrani, and G. Barca, Phys. Rev. D 99, 063534, (2019).
	
	\bibitem{barca} G. Barca, P. di Antonio, G. Montani, A. Patti, Phys. Rev. D 99, 123509, (2019).
	
	\bibitem{40} B. F. Schutz, Jr., Phys, Rev. D 4, 3559, (1971)
	
	\bibitem{50} P. A. M. Dirac,  Proc. Roy Soc. (London) A395,1, (1968).
\end{thebibliography}
\end{document}